\documentclass[twocolumn,showpacs,preprintnumbers,amsmath,amssymb]{revtex4}
\usepackage{latexsym}
\usepackage{graphicx}
\usepackage{dcolumn}
\usepackage{bm}
\usepackage{amsmath}
\usepackage{hyperref}
\begin{document}
\title{Localized modes in defective multilayer structures}
\author{S. Roshan Entezar$^{1}$}
\author{A. Namdar$^2$}

\affiliation{$^1$Physics faculty,  University of Tabriz , Tabriz,
Iran} \affiliation{ $^2$Physics Department, Azarbaijan University of
Tarbiat moallem, Tabriz, Iran}%
\date{\today}
\begin{abstract}
In this paper, the localized surface modes in a
defective multilayer structure has been investigated. It is shown
that the defective multilayer structures can support two different kind of localized modes
depending on the position and the thickness of the defect layer. One
of these modes is localized at the interface between the multilayer
structure and a homogeneous medium (the so-called surface mode) and
the other one is localized at the defect layer (defect localized
mode). We reveal that the presence of defect layer
pushes the dispersion curve of surface modes to the lower or the
upper edge of the photonic bandgap depending on the homogeneous
medium is a left-handed or right-handed medium (e.g. vacuum), respectively. So, the existence
region of the surface modes restricted. Moreover, the effect of defect on the energy flow velocity of
the surface modes is discussed.
\end{abstract}
\maketitle
\section{Introduction}
The advent of photonic crystal (PC) materials \cite{Photonic Band
Gaps and Localization, Photonic Crystals:} excited a lot of interest
toward the existence of surface modes (SMs) at the interfaces of
such materials \cite{W. M. Robertson, R. D. Meade, J. M. Elson, F.
Ramos-Mendieta}. These modes remain localized around
the PC surface in a manner similar to surface plasmons on metal
slabs \cite{N. W. Ashcroft}. In real materials, PCs are always
finite in size and, therefore, SMs can exist \cite{J. M. Elson, F.
Ramos-Mendieta, F.
Ramos-Mendieta1, P. Etchegoin}. The existence
of SMs can directly affect the performance and efficiency of PCs in
applications. Thus, it is important to study the SMs in PCs. The
first experimental observation of such SMs came by Robertson et al
\cite{W. M. Robertson}. In the experiment, the authors employed a
standard attenuated total reflection (ATR) setup, widely used for
surface-plasmon observations on metals \cite{A. Otto, A. Otto1}. The
majority of the subsequent theoretical studies focused on the
existence of such SMs in various PC structures. It was found that
the frequency and in many cases even the mere existence of SMs are
strongly influenced by the way the periodic PC is terminated \cite{R.
D. Meade, F. Ramos-Mendieta}. Nevertheless, the initial acute
interest for SMs had somewhat subsided until recently. The need to
understand and engineer SMs came back to light, when these deemed to
play a key role in newly discovered PC phenomena. In particular, it
was found that SMs can strongly influence the subdiffraction
focusing properties of PC based slab superlenses \cite{R. Moussa, C.
Luo, S. Xiao, P. Kramper}. Moreover, coupling to such SMs in PC
subwavelength-width waveguides leads to a highly directional exit
beam \cite{Esteban Moreno, I. Bulu}. Despite the intensive research
on PC surface phenomena \cite{W. M. Robertson, R. D. Meade, F.
Ramos-Mendieta, J. M. Elson}, one aspect of the SM propagation
remains unexplored. When the periodicity of PC is broken by
introducing a defect into a PC, a defect mode will appear
inside the photonic bandgap (PBG) due to change of the interference
behavior of light, whose properties would be determined by the
nature of the defect. The introduction of defect layers in
one-dimensional PCs can create defect modes within the
PBGs, just as defect layers in semiconductor superlattices may
result in electron defect states in the band gaps. A natural
question arises: what is the effect of defect on the SMs?

In this paper, we investigate the effect of defect on the SMs at the
interface between a multilayer structure and a homogeneous
medium (the interface of the structure). We show that
two different kind of localized modes can be created in the defective multilayer structure
depending on the position and the thickness of defect layer. One
of these modes is a SM and has a peak intensity at the interface of the structure. The other one is localized at
the defect layer which we call it the defect localized
mode (DLM). Furthermore, it is shown that the existence regions of
the surface modes and defect localized modes depends on
the position and thicknesses of the defect layer. As well, we show that
the energy flow velocity of
the surface modes has not been considerably affected due to the existence of the defect layer in the periodic multilayer
structure. However, the energy flow velocity of the defect localized
modes is affected by the position and the thickness of the defect layer. In Sec. II, we introduce the
model of the system under consideration. In Sec. III, the properties
of localized modes are studied. Finally, Sec. IV concludes with brief comments.

\section{Formalism}

We wish to describe localized modes that form at the interface between a homogeneous medium
of low refractive index, $n_0=\sqrt{\varepsilon_0 \mu_0}$, and a
defective multilayer structure with layers of refractive indices
$n_1=\sqrt{\varepsilon_1 \mu_1}$ and $n_2=\sqrt{\varepsilon_2
\mu_2}$ and thicknesses $d_1$ and $d_2$. Here we adopt the typical
values used in reference \cite{A. Namdar};  $\varepsilon_0=-1$, $
\mu_0=-1$, $\varepsilon_1=4$, $ \mu_1=1$, $\varepsilon_2=2.25$, $
\mu_2=1$, $d_1=1~cm$ and $d_2=1.65~cm$. We suppose that the defect layer
with thickness $d_d$ and refractive index $n_1$ located after $\eta,~(\eta=0,1,2,...)$
complete periods ($d=d_1+d_2$) of multilayer structure, hence $\eta$ describes
the position of the defect layer in the structure. We choose a coordinate
system in which the layers have normal vector along $OZ$ and we
consider the propagation of monochromatic TE-polarized waves
described by \cite{Morozov2004}
\begin{eqnarray}
\textbf{E}&=&E_y(z)\hat{e}_y e^{i(k\beta x-\omega t)},\\\nonumber
\textbf{H}&=&(H_x(z)\hat{e}_x+H_z(z)\hat{e}_z)e^{i(k\beta x-\omega
t)}, \label{eq3}
\end{eqnarray}
with the electric field E in the y direction. Here, $\omega$ is the
angular frequency, $k = \omega/c$ is the vacuum wavenumber, $\beta=\frac{k_x}{k}$ and
$k_x$ is the x-component of the wave-vector of modulus $k(z) =
k n(z)$. We look for stationary solutions propagating along
the interface which satisfy the following scalar Helmholtz-type
equation
\begin{equation}
\left[\frac{d^2}{dz^2}-k_x^2+\frac{\omega^2}{c^2}\varepsilon(z)\mu(z)-\frac{1}
{\mu(z)}\frac{d\mu}{dz}\frac{d}{dz}\right]E_y=0.\label{eq4}
\end{equation}
In the periodic structure, the waves are the Bloch modes
$E_y(z)=\psi(z)e^{i K_b z}$ where $K_b$ is the Bloch wavenumber, and
$\psi(z)$ is the Bloch function, which is periodic with the period
of the photonic structure (see Ref. \cite{Yariv1984}). In the
periodic structure the waves will be decaying provided $K_b$ is
complex; and this condition defines the spectral band gaps of an
infinite multilayer structure. For the calculation of the Bloch
modes, we use the well-known transfer matrix method
\cite{Yariv1984}. To find the localized modes, we take solutions of Eq.
(\ref{eq4}) in a homogeneous medium and the Bloch modes in the
periodic structure and satisfy the conditions of continuity of the
tangential components of the electric and magnetic fields the interface of the structure
\cite{Martorell2006}. In this way, we can obtain the exact
dispersion relation $k = k(\beta)$ for TE-polarized localized modes by
numerically solving the following dispersion condition:
\begin{widetext}
\begin{equation}
\frac{q_0\mu_1}{k_1\mu_0}=-i\frac{\{(A+B)U_{\eta-1}-U_{\eta-2}\}(\lambda-A)-
\{(A+B)^\star
U_{\eta-1}-U_{\eta-2}\}\tilde{B}}{\{(A-B)U_{\eta-1}-U_{\eta-2}\}(\lambda-A)+
\{(A-B)^\star U_{\eta-1}-U_{\eta-2}\}\tilde{B}}\label{dispersion}
\end{equation}
\end{widetext}
where $q_0=k\sqrt{\beta^2-n_0^2}$ and $k_1=k\sqrt{n_1^2-\beta^2}$.
Here, $A,~B$ are the elements of the transfer matrix of the
multilayer structure, $\lambda$ is the eigenvalue of the transfer
matrix $\left[\begin{array}{cc} A & B \\  B^\ast & A^\ast \\ \end{array}\right]$ \cite{Morozov2004},
 $\tilde{B}=B e^{-2ik_1d_d}$ and $U_{\eta} =\sin{((\eta+1)K_b
d)}/sin{(K_b d)}$.

\section{Numerical results and discussions}

\begin{figure}[b]
\includegraphics[width=7cm]{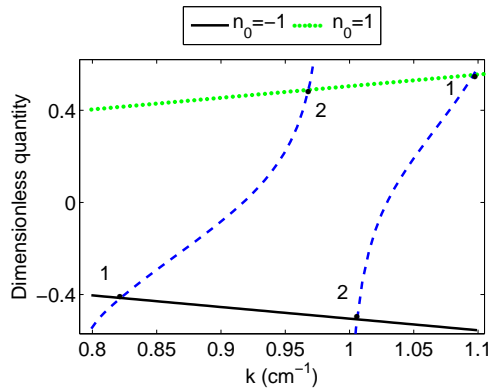}
\caption{\label{Figure1} Graphical representation of the solutions of
dispersion condition (Eq. (\ref{dispersion})). Here the solid and dotted lines
show the left-hand side of Eq. (\ref{dispersion}) for the cases $n_0=-1$ and $n_0=1$ respectively,
while the dashed lines represent the right-hand side of Eq. (\ref{dispersion}). In our calculations we take
the following values:
$d_d=0.8~d_1$, $\eta=3$ and $\beta=1.14~cm^{-1}$.}
\end{figure}
\begin{figure}[t]
\includegraphics[width=9cm]{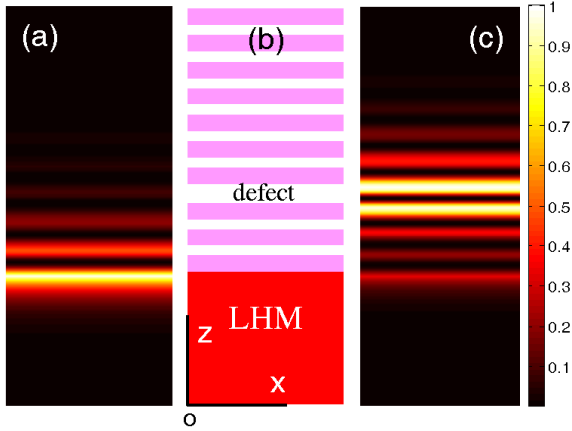}
\caption{\label{Figure2} Geometry and examples of the surface and defect modes:
(a) SM, $k=0.821~cm^{-1}$, (b) geometry and (c) DLM,
$k=1.019~cm^{-1}$. These modes are
corresponding to the points $1$ and $2$ in Fig. \ref{Figure1},
respectively}. Here, $\beta =1.14$ and $n_0=-1$.
\end{figure}

In the following, we want to discuss the effect of defect layer on
the dispersion properties of the localized modes. In the used
structure, we assumed that the homogeneous medium is a left-handed metamaterial (LHM)
with $n_0=-1$. For comparison, we also investigated the dispersion of
the corresponding localized modes in the structure, where the
homogeneous medium is replaced by vacuum ($n_0=1$). To do this, we plotted
two sides of the dispersion condition (Eq. (\ref{dispersion})) vs. $k$ in
the first spectral bandgap in Fig. \ref{Figure1}. In this figure the solid and dotted lines
show the left-hand side of Eq. (\ref{dispersion}) for the cases $n_0=-1$ and $n_0=1$ respectively,
while the dashed lines represent the right-hand side of Eq. (\ref{dispersion}).
Here we take the following values: $d_d=0.8~d_1$, $\eta=3$ and $\beta=1.14~cm^{-1}$. Unlike the
defectless structure \cite{A. Namdar}, we see that the dispersion
condition has two solutions (points $1$ and $2$ in Fig.
\ref{Figure1}). So, there are two different modes for a given
$\beta$. To describe these modes, we plotted the intensity
distribution of these modes as a function of $z$ in Fig.
\ref{Figure2}. As one can see, both of these modes are localized
modes. But, the mode $1$ is localized the interface of the structure (SM) (see Fig.
\ref{Figure2}(a)), while the mode $2$ is localized at the defect
layer (DLM) (see Fig. \ref{Figure2}(c)).

\begin{figure}[b]
\includegraphics[width=9cm]{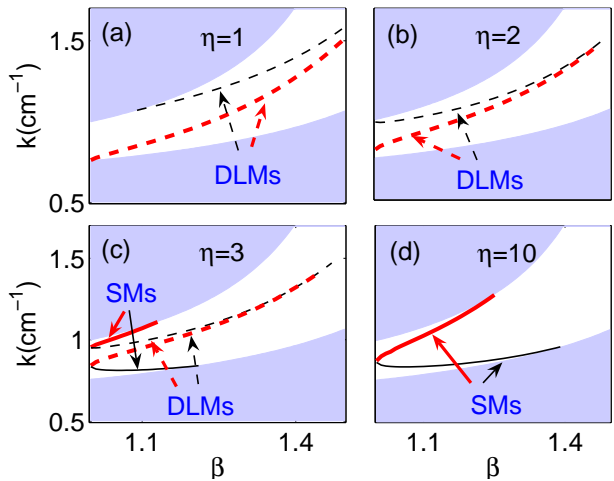}
\caption{\label{Figure3} The effect of defect position on the
dispersion properties of SMs (solid lines) and DLMs (dashed lines) in the first band gap. Here,
the thin lines show the case for $n_0=-1$ and the thick lines show the
case for $n_0=1$. The shaded regions show propagation bands of the
multilayer structure. Here, $d_d=0.8~d_1$.}
\end{figure}
Our investigations indicate that the existence region of the SMs and
the DLMs strongly depends on the position of defect layer. To show this,
we plotted the dispersion curves of the localized modes in the first
spectral gap as a function of $\beta$ for different defect position
in Fig. \ref{Figure3}. As one can see from Figs.
\ref{Figure3}(a),(b), when the position of the defect layer approaches to the surface of the structure (small $\eta$),
the coupling between the incident wave and the defect mode strongly take places and leads to the formation of the DLM. But
by increasing $\eta$,
the interaction between the incident wave and defect mode
becomes weaker (see Figs.
\ref{Figure3}(c)), so that in the case of sufficiently big $\eta$
the DLM disappears in the benefit of the appearance of the SM.
This case ($\eta~\gg$) corresponds to the defectless structure (see Figs.
\ref{Figure3}(d)).

In Fig. \ref{Figure3} we also studied the dispersion property of two different homogeneous media (i.e. LHM and vacuum).
We see that when the homogeneous medium is a LHM ($n_0=-1$),
the dispersion cure of the SMs settle in the near by the lower edge of the photonic bandgap and finally
disappears by decreasing the distance of the defect layer from the interface of the structure. This is in
contrast to the case where the LHM is replaced by the
vacuum. For this case the dispersion cure of the SMs moves toward the upper edge of the photonic bandgap.
\begin{figure}[t]
\includegraphics[width=9cm]{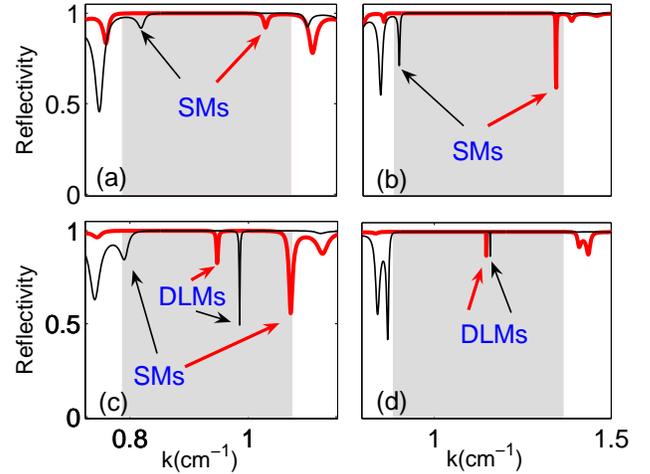}
\caption{\label{Figure4} The calculated ATR spectrum of a defectless
multilayer structures with (a) $\beta=1.09$, (b) $\beta=1.29$ and a
defective multilayer structures with (c) $\beta=1.09$, (d)
$\beta=1.29$. Here, the thin lines show the case for $n_0=-1$ and the thick lines show the
case for $n_0=1$. The other parameters
are $d_d=0.8~d_1$, $\eta=3$ and $\acute{n}=3$.}
\end{figure}
\begin{figure}[t]
\includegraphics[width=7cm]{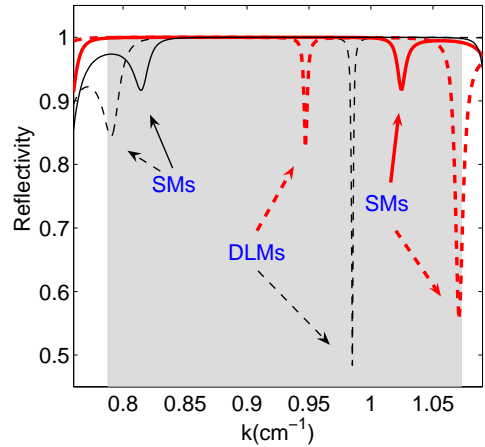}
\caption{\label{Figure5} The calculated ATR spectrum of a defective
multilayer structures for different width of the defect layer
$d_d=0.1~d_1$ (solid line), $d_d=0.8~d_1$ (dashed line).
Here, the thin lines show the case for $n_0=-1$ and the thick lines show the
case for $n_0=1$. The other parameters
are $\beta=1.09$, $\eta=3$ and $\acute{n}=3$.}
\end{figure}

As stated in Fig. \ref{Figure2}, the localized modes have evanescent
nature and due to this evanescent nature, they will not interact
directly with an incoming planewave. So, they can be excited by the
ATR method. This technique has previously been invoked for the
investigation of various types of surface polaritons, e.g.,
plasmon-polaritons in metals \cite{A. Otto, A. Otto1},
phonon-polaritons in ionic crystals \cite{R. Ruppin, V.V. Bryksin},
exciton-polaritons in semiconductors \cite{I. Hirabayashi, J.
Lagois} and magnon-polaritons in magnetic materials \cite{J.
Matsuura, M. Fukui}. We consider the ATR geometry shown in Fig. 1 in
reference \cite{A. Namdar1}. Here, the uniform medium represents a
gap layer of the width L that separates dielectric and layered
structure. For an incident angle larger than the angle of the total
internal reflection, the electromagnetic field incident from an
optically dense medium (dielectric) with refractive index $\acute{n}>n_0$ will penetrate
the gap as an evanescent wave, which can interact with the
evanescent localized modes. We have calculated the reflectivity of
the ATR geometry, using classical electromagnetic theory. A
calculated ATR spectrum for the cases of defectless and defective
multilayer structures are shown in Fig. \ref{Figure4} for two
different incident angle (or two different
$\beta=\acute{n}sin(\theta)$). Here, the shaded regions show the
first spectral bandgap of the multilayer structure. In Figs. \ref{Figure4}(a),(b)) there is only a deep in the ATR spectrum in the
spectral bandgap corresponding to a SM.
However, in Fig. \ref{Figure4}(c) we reveal two sharp deeps in the ATR spectrum corresponding to a DLM and a SM. In Fig.
\ref{Figure4}(d), we can see only a deep in the spectral bandgap
related to a DLM.
\begin{figure}[b]
\includegraphics[width=9.cm]{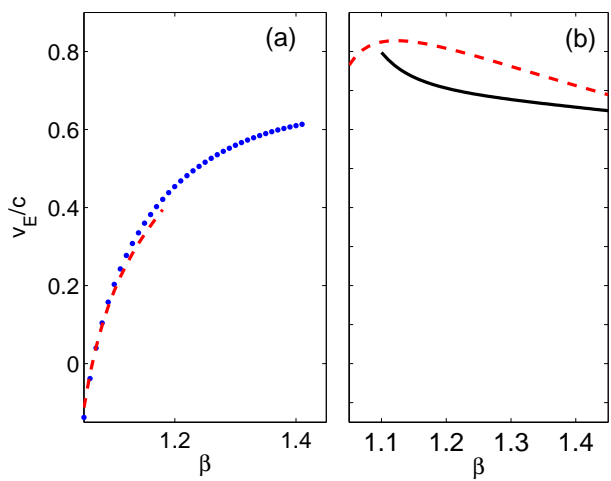}
\caption{\label{Figure6} The energy flow velocity of (a) SMs  and
(b) DLMs for different positions of the defect layer, $\eta=1$
(solid lines), $\eta=3$ (dashed lines) and $\eta=7$ (dotted lines).
Here, $d_d=0.8~d_1$ and $n_0=-1$.}
\end{figure}
\begin{figure}[t]
\includegraphics[width=9.cm]{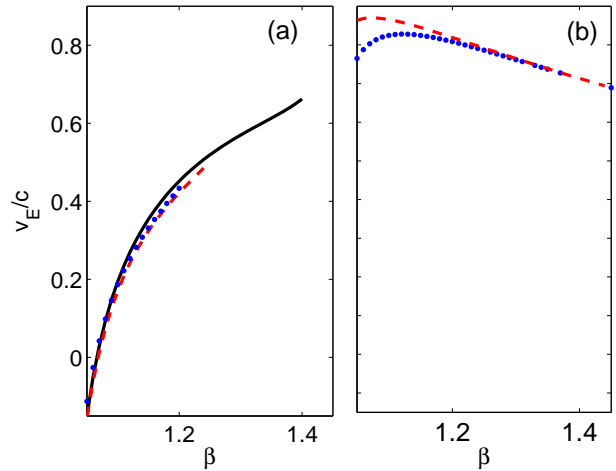}
\caption{\label{Figure7} The energy flow velocity of (a) SMs  and
(b) DLMs for different thicknesses of the defect layer,
$d_d=0.1~d_1$ (solid lines), $d_d=0.5~d_1$ (dotted lines) and
$d_d=0.8~d_1$ (dashed lines). Here,  $\eta=3$ and $n_0=-1$.}
\end{figure}
Our investigations reveal that the existence region for the SMs and
the DLMs not only depends on the position of the defect layer ($\eta$)
but also it depends on the thickness of defect layer ($d_d$). To
show this, we plotted the reflectivity of the ATR geometry as a
function of frequency ($k=\frac{\omega}{c}$) for different
thicknesses of defect layer with a fix position in Fig. \ref{Figure5}.
It is seen that for a thin defect layer, the multilayer structure
behaves like as a defectless structure (see solid lines in Fig. \ref{Figure5}) \cite{A. Namdar}. Increasing the thickness
of the defect layer $d_d$, will push the SM to the lower (upper) edge of the bandgap in the case of $n_0=-1$ ($n_0=1$)
then it will disappear for the sufficiently thick defect layer, and the DLM will appear within the bandgap.
In the other word, for a big $d_d$ the structure loses the ability to support
the SMs. Therefore, the modes
localize at the defect layer.

Finally, it is interest of us to know the effect of defect layer on the energy flow velocity of the localized
modes. It is well known that the energy flow velocity $v_E$ relates total energy flow
${\langle S\rangle}$ to the energy density integrated over $z$ as
$v_E=\frac{\langle S\rangle}{\langle U\rangle}$. We demonstrated the
effect of defect position and thickness on the energy flow velocity
of localized modes in the defective multilayer structure in Figs.
\ref{Figure6} and \ref{Figure7}, respectively.
It is seen that in the presence of uniform LHM, the energy flow velocity of the SMs can be positive or
negative (see Figs. \ref{Figure6}(a) and \ref{Figure7}(a)), which
indicates the considered structure has the potential to support forward and
backward energy flow (this is the noticeable advantage of using LHM). On
the other hand, Figs. \ref{Figure6}(b) and \ref{Figure7}(b) show
that the DLMs have only positive energy flow velocity. Accordingly,
in the used structure the presence of LHM can not
affect the direction of energy follow of the DLM. In addition, Figs.
\ref{Figure6}(a) and \ref{Figure7}(a) reveal that the presence of
defect layer can not affect considerably the energy flow velocity of
the SMs, whereas the energy flow velocity of the DLMs depends on
the position and thickness of defect layer (see Figs.
\ref{Figure6}(b) and \ref{Figure7}(b)).

\section{Conclusion}

Briefly, we have studied the dispersion characteristics of the localized modes
in the presence of defect layer in the periodic multilayer structures. It was shown
that the structure can support both the SMs and the DLMs which localize at the interface
of the structure and the position of the defect, respectively.
Moreover, we have revealed that the existence regions of the SMs and the DLMs
depend on the position and thickness of the defect layer. We show that
the dispersion curves of the SMs settle near by the lower (upper) edge of the photonic bandgap
for the case of the uniform medium is a LHM (vacuum), and finally
disappear by decreasing the distance of the defect layer from the surface of the structure. As well, we have shown that
the existence of the defect in the structure has not a considerable effect on the energy flow velocity of
the SMs.

\section{Acknowledgment}

The authors thank Prof. Yuri S. Kivshar for valuable suggestions and useful guidance.

\bibliography{apssamp}
{}
\end{document}